\documentclass[%
 aip,
 apl,%
 amsmath,amssymb,
reprint,%
]{revtex4-1}

\usepackage{graphicx}
\usepackage{dcolumn}
\usepackage{bm}

\newcommand{\IPCMS}{Institut de Physique et Chimie des Mat\'{e}riaux de Strasbourg, UMR 7504 CNRS, Universit\'{e} de Strasbourg, 23 rue du Loess, BP 43, 67034 Strasbourg Cedex 2, France}
\newcommand{\UMPhy}{Unit\'{e} Mixte de Physique CNRS, Thales, Univ. Paris-Sud, Universit\'{e} Paris-Saclay, 91767 Palaiseau, France}

\newcommand{\IEF}{Centre de Nanosciences et de Nanotechnologies, CNRS, Univ. Paris-Sud, Universit\'e Paris-Saclay, 91120 Palaiseau, France}

\begin{document}


\title{Determining key spin-orbitronic parameters by means of propagating spin waves}

\author{O. Gladii}
\affiliation{\IPCMS}
\author{M. Collet}
\affiliation{\UMPhy}
\author{Y. Henry}
\affiliation{\IPCMS}
\author{J.-V. Kim}
\affiliation{\IEF}
\author{A. Anane}
\affiliation{\UMPhy}
\author{M. Bailleul}
\affiliation{\IPCMS}

\date{\today}

\begin{abstract}

We characterize spin wave propagation and its modification by an electrical current in Permalloy(Py)/Pt bilayers with Py thickness between 4 and 20 nm. First, we analyze the frequency non-reciprocity of surface spin waves and extract from it the interfacial Dzyaloshinskii-Moriya interaction constant $D_s$ accounting for an additional contribution due to asymmetric surface anisotropies. Second, we measure the spin-wave relaxation rate and deduce from it the Py/Pt spin mixing conductance $g^{\uparrow\downarrow}_{eff}$.  Last, applying a \textit{dc} electrical current, we extract the spin Hall conductivity $\sigma_{SH}$ from the change of spin wave relaxation rate due to the spin-Hall spin transfer torque. We obtain a consistent picture of the spin wave propagation data for different film thicknesses using a single set of parameters $D_s=0.25$~pJ/m, $g^{\uparrow\downarrow}_{eff} = 3.2\times 10^{19}$ m$^{-2}$ and $\sigma_{SH}=4\times10^{5}$~S/m.
\end{abstract}

\maketitle
\section{Introduction}

The control of magnetization dynamics is a key aspect in the development of spintronic devices. It can be implemented conveniently by exploiting spin-orbit effects in heavy metal (HM)/ferromagnet (FM) structures. The strong spin orbit coupling in the heavy metal leads to a modification of the magnetization dynamics in the adjacent ferromagnet which can be taken advantage of in different manners. On one hand, the breaking of the inversion symmetry at the interface gives rise to an antisymmetric exchange interaction~\cite{Dzialoshinskii} known as interfacial Dzyaloshinskii-Moriya interaction (iDMI). Because this chiral interaction favors non-uniform magnetic textures, it stabilizes magnetic vortices and, in particular, magnetic skyrmions, which have potential for diverse applications such as data storage \cite{Kiselev_2011}, microwave detection and energy harvesting \cite{Finocchio2015}, probabilistic computing \cite{Pinna2018}, and reservoir computing \cite{Prychynenko2018}. On the other hand, spin-Hall effect spin-transfer torque (SHE-STT)  can be used for electrical control of magnetization dynamics, including damping modulations~\cite{Kasai2014,Demidov2011a}, magnetization switching~\cite{Liu2012, Miron2011} or self-induced oscillations~\cite{Collet2016,Demidov2012}. In HM/FM systems, STT arises from the transfer of angular momentum from the spin current ejected from HM to the local magnetization of FM, the spin current in HM being usually generated by the spin Hall effect (SHE)~\cite{Dyakonov1971,Sinova2015}. The efficiency of SHE-induced STT is determined by the ability of the electric field to be converted into the transverse spin current, which is quantified via the spin Hall conductivity, and the ability of the spin current to penetrate through the HM/FM interface. Despite the difficulty that improvements in the SHE-STT efficiency can be accompanied by a degradation of the zero current damping due to spin pumping~\cite{Saitoh2006,Tserkovnyak2002}, a number of experiments have demonstrated the possibility to significantly reduce or even completely compensate the damping.\cite{Hamadeh2014,Collet2016}  
In current-induced domain-wall motion experiments, the two effects combine with each other particularly favorably, the iDMI stabilizing domain-walls of the Neel type for which the SHE-STT is particularly efficient~\cite{Emori2013}. Interestingly, the two effects are also expected to  coexist in another type of non-unform magnetization dynamics, namely the propagation of spin waves. Indeed, in the so-called Magnetostatic Surface Wave geometry (equilibrium magnetization oriented in the film plane, perpendicular to the SW wave-vector), the iDMI translates directly into a spin-wave frequency non-reciprocity, \textit{i.e.} a difference of frequency between counterpropagating SW,\cite{Belmeguenai2015,Moon2013,Cortes-Ortuno2013,Kostylev2014} and the SHE STT translates into a current-induced modification of the spin-wave relaxation rate\cite{Demidov2014,Gladii2016a}. However, up to now, the two effects could not be observed simultaneously, iDMI being extracted in ultrathin films for which this interfacial effects has a very large impact but for which spin waves do not propagate far enough to determine their relaxation rate.

In this work, we consider Py/Pt bilayers with varying Py thickness and investigate systematically how the propagation of spin waves is influenced by the iDMI interaction, the spin pumping and the spin Hall effect. For this purpose, we employ the technique of propagating spin-wave spectroscopy~\cite{Vlaminck2010,Haidar2013,Gladii2016a} which allows one to determine precisely the frequency and the relaxation rate of spin waves, together with their current-induced modification. First, the strength of Dzyaloshinskii-Moriya interaction is estimated from the frequency shift between two counter-propagating spin waves taking into account the additional contribution to the frequency shift arising from magnetic asymmetry which becomes sizeable for the thicker films\cite{Gladii2016}. Then, the spin wave relaxation rate is measured with and without applied dc current to examine the two reciprocal effects, namely the increase of the spin-wave relaxation rate due to spin pumping into the HM layer, and its modulation by an electrical current mediated by the SHE-STT. The results for different films thicknesses are accounted for using a single set of spin-orbitronic parameters, which provides us with a clear global picture of the spin-orbit related phenomena in this system.

\begin{figure}
\includegraphics[width=8.2cm]{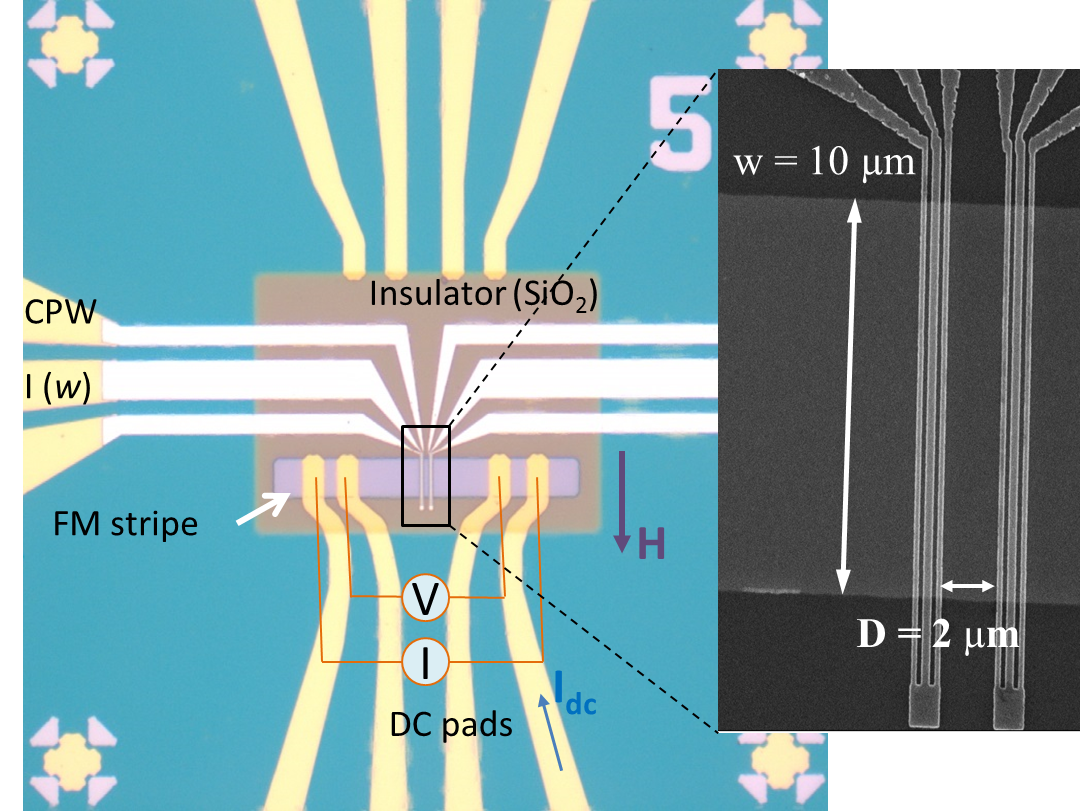}
\caption{Optical microscopy image of a spin wave device consisting of a Py/Pt strip, the coplanar waveguide connected to a pair of microwave antenna and dc pads.} 
\label{device}
\end{figure}

\section{Description of the experiment}

A set of Ti(5 nm)/Py($t$)/Pt(5,10 nm) films with Py thickness $t = 4, 7, 10, 20$~nm have been grown, together with a Ti(5)/Py(4)/Ti(5) film. The latter is a reference stack where the spin-orbit interaction is negligible. All films have been deposited by magnetron sputtering on intrinsic silicon substrates with a thermal oxide layer of about 100 nm. Then, the experimental devices have been fabricated using standard lithography processes as described in Refs.~[\onlinecite{Vlaminck2010,PhD_Gladii}] (Fig.~\ref{device}). Each device consists of a ferromagnetic strip of width $w$, two microwave antennas patterned on top and connected to coplanar waveguides (CPW), and DC pads attached to the strip for injecting a \textit{dc} current.The FM/HM strip is covered by an insulating SiO$_2$ layer to avoid electrical contact with antenna. Measurements proceed as follows: an external magnetic field $H$ is applied along the axis of the antennas to orient the equilibrium magnetization in the so-called magnetostatic surface wave configuration, the spin wave is excited by the microwave current circulating in one antenna and is detected by measuring the magnetic flux induced in the other antenna. Each of the antennas can work either as an emitter or receiver, which allows one to generate and detect counter-propagating spin waves. The measured mutual-inductance $\Delta L_{ij}$, where $i$ and $j$ ($=1,2$) correspond to receiving and emitting antenna respectively, gives the information about the characteristics of propagating spin waves. For the given antenna geometry the wave vector of the excited SW is $k=7$~rad$/\mu$m$^{-1}$, which is defined by the Fourier transform of the spatial distribution of the microwave current. To characterize the SW spatial exponential decay, we fabricated devices with various distances $D$ between antennas ($D=0.5$ to 3 $\mu$m edge to edge). 

\begin{figure}
\includegraphics[width=8.2cm]{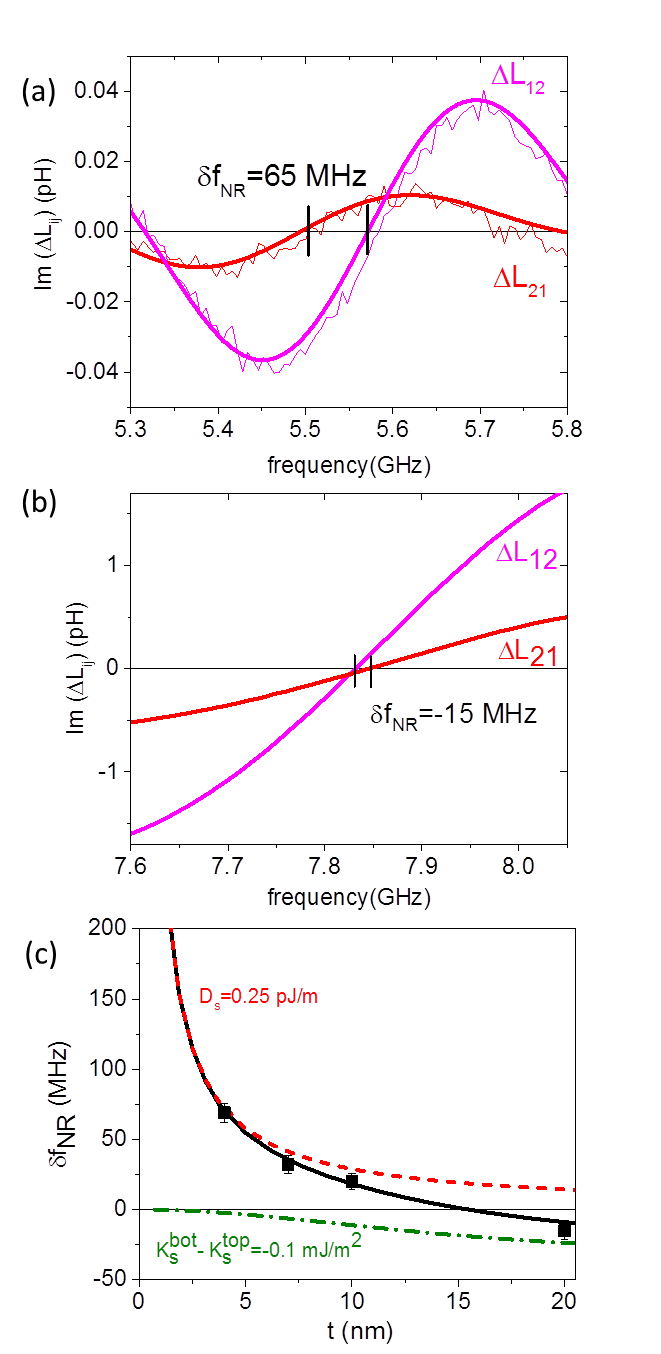}
\caption{Imaginary part of the mutual-inductance spectra measured at $\mu_0H=37$~mT in Py(4)/Pt (a) and Py(20)/Pt (b) devices with $D=1$~$\mu$m for spin waves with $k<0$ and $k>0$. (c) Frequency non-reciprocity measured in Py/Pt bilayers (black squares) for $\mu_0H=37$~mT . Dashed line and dashed-dotted line are the frequency non-reciprocity induced by iDMI and magnetic asymmetry, respectively, calculated for $D_S=0.25$~pJ/m and $\Delta K=-0.1$~mJ/m$^2$. The sum of the two contributions is shown as a solid line.} 
\label{DMI}
\end{figure}

\section{Interfacial Dzyaloshinskii-Moriya interaction}

The influence of iDMI on propagating surface spin waves is studied by measuring the frequency shift between counter-propagating spin waves in Py/Pt bilayers with varying Py thickness. The measurements are performed at zero dc current. Fig.~\ref{DMI}(a) and (b) show the mutual inductance signals $\Delta L_{12}$ and $\Delta L_{21}$ measured at $\mu_0H_0=37$~mT in the Py(4)/Pt and Py(20)/Pt films, respectively. $\Delta L_{12}$ and $\Delta L_{21}$ correspond to the waves with $k<0$ and $k>0$, respectively. As one can see, the frequency of the left-moving spin wave is shifted with respect to the frequency of the right-moving one with a difference defined as $\delta f_{NR}=f_{12}-f_{21}$ (here $f_{12}$ and $f_{21}$ are the frequencies at which the curves intercept the horizontal axis). The values for $t=4$~nm and $t=20$~nm are $\delta f_{NR}=65$~MHz and $\delta f_{NR}=-15$~MHz, respectively, while in the reference Py(4)/Ti(5) stack, $\delta f_{NR}$ is as small as 0.8 MHz (not shown). In Fig.~\ref{DMI} (c) the frequency non-reciprocity is plotted as a function of Py thickness (symbols). 

The frequency shift induced by iDMI is expected to vary as~\cite{Belmeguenai2015,Di2015} 
\begin{equation}
\delta f_{NR,iDMI}=\frac{2\gamma D_s}{\pi M_st}k,
\label{eqDMI}
\end{equation}
where $D_s$ is the thickness independent iDMI constant. From Eq.~\ref{eqDMI}, the frequency shift is expected to be linear and odd in $k$ and inversely proportional to the film thickness, which is the standard scaling for a purely interfacial effect. However, we observe in Fig.~\ref{DMI}(c) a deviation from the $1/t$ law for thicker films and even a change of sign for $t>15$~nm, where the positive sign of $\delta f_{NR}$ indicates a higher frequency for the wave traveling in negative direction, whereas the negative sign indicates a higher frequency for the wave moving in positive direction. 

This unusual thickness dependence is attributed to a magnetic asymmetry across the film thickness, which produces a frequency non-reciprocity opposite to that induced by the iDMI in our case. As discussed in Ref.~[\onlinecite{Gladii2016,Henry2016}], such frequency shift results from the dipolar tendency of magnetostatic surface wave to localize more on one surface or the other depending on its propagation direction. In our Ti/Py/Pt trilayers, we attribute the magnetic asymmetry to a difference between the surface anisotropies at the top ($K_s^{top}$) and bottom ($K_s^{bot}$) interfaces. The expression of $\delta f_{NR}$ derived for this situation is:\cite{Gladii2016}  
\begin{equation}
\delta f_{NR,Ks}\simeq\frac{8\gamma}{\pi^3}\frac{K_s^{bot}-K_s^{top}}{M_s}\frac{k}{1+\frac{\Lambda^2\pi^2}{t^2}},
\label{eqKs}
\end{equation}
where $\Lambda$ is the exchange length and $\frac{\Lambda^2\pi^2}{t^2}$ is the splitting between the fundamental mode and the first order perpendicular standing spin wave. As follows from Eq.~\ref{eqKs}, this frequency shift is linear and odd in $k$ as in the case of iDMI-induced shift, but it shows a nearly quadratic dependence on the film thickness instead of the $1/t$ dependence appropriate for iDMI.  This distinct thickness dependence is actually the only criterion to separate the contributions of both effects, which obey the same symmetry (odd in $k$, but also odd in $H$)~\cite{Gladii2016}. 

The best fit was obtained by summing the two contributions Eqs.~\ref{eqDMI} and \ref{eqKs} with an iDMI constant $D_s=0.25$~pJ/m and a difference of surface anisotropy $K_s^{bot}-K_s^{top} = -0.1$~mJ/m$^2$ [solid line in Fig.~\ref{DMI} (c)]. To confirm these analytical findings, we also performed a rigorous numerical calculation ~\cite{Henry2016}, subdividing the film into slabs of thickness $0.125$~nm and introducing the effective fields corresponding to iDMI and uniaxial magnetic anisotropy only in  the top cell. The simulated frequency non-reciprocity is exactly the same as the one derived from the analytical treatment given above. In Fig.~\ref{DMI}(c) the individual contributions from the iDMI and from the difference in surface anisotropies are plotted as a red dashed line and a green dot-dashed line, respectively. In very thin films the dominant contribution is that of the iDMI, which is consistent with the purely interfacial nature of this effect, while the contribution of the difference in surface anisotropies is negligible. On the contrary, in moderately thin film the influence of the magnetic asymmetry becomes more important: it fully compensates the iDMI-induced frequency shift at $t=15$ nm and becomes dominant in thicker films. A similar value of iDMI constant for Py/Pt interface was reported in Ref.[~\onlinecite{Nembach2015}], where the frequency non-reciprocity was studied using Brillouin light scattering (BLS). Moreover, a similar thickness dependence of $f_{NR}$ was reported in Ref.~\onlinecite{Stashkevich2015}, in which a very small frequency non-reciprocity was calculated for a Py thickness of 10 nm. Due to the limited frequency resolution of the conventional BLS technique ($\pm75$ MHz), this frequency shift could not be resolved. By using PSWS with a resolution of a few MHz, we can clearly follow the frequency non-reciprocity over an extended thickness range and distinguish the contributions from iDMI and from the difference between the surface anisotropies at both interfaces.

\begin{figure}
\includegraphics[width=8.2cm]{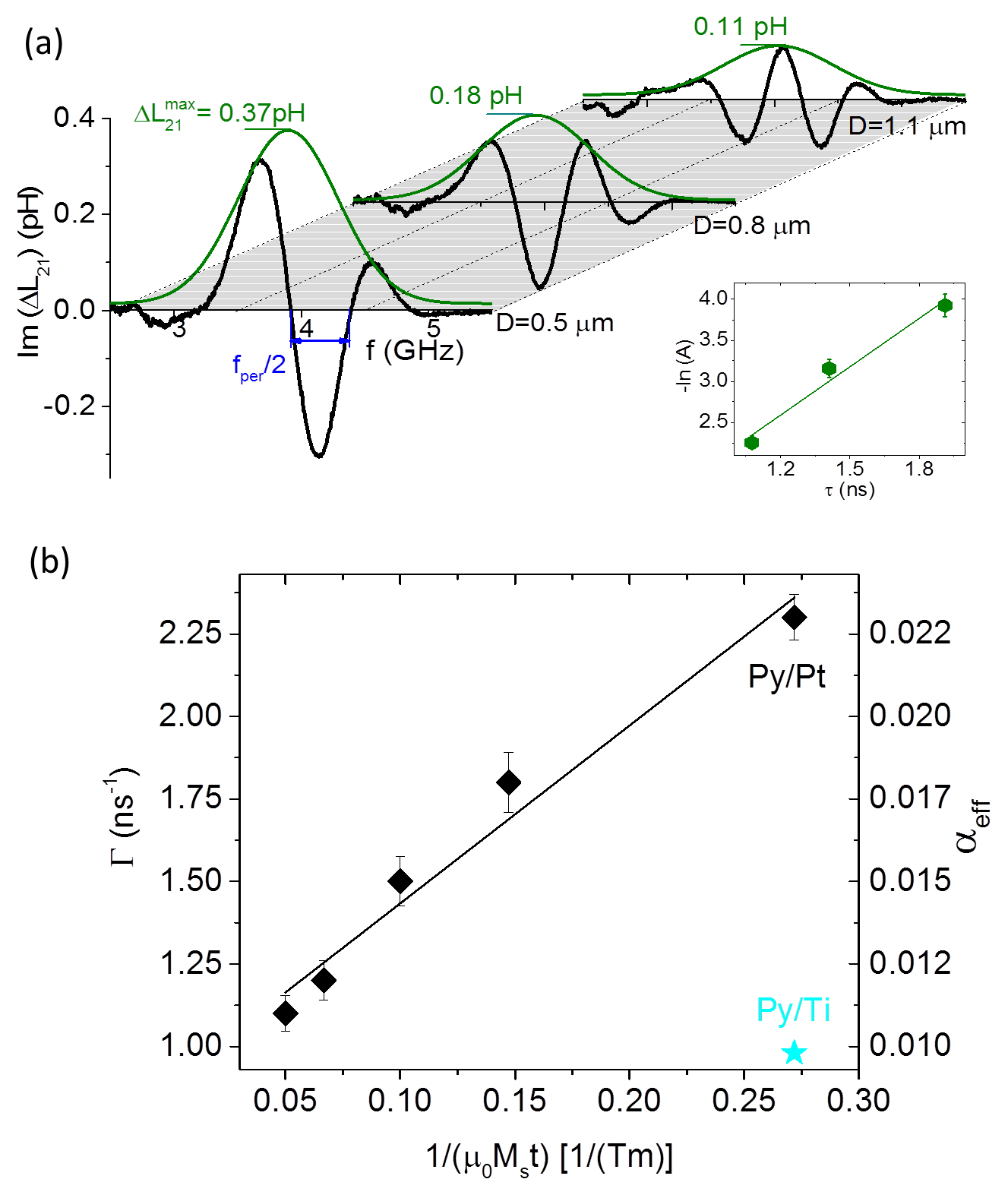}
\caption{(a) Imaginary part of the mutual-inductance measured in Py(4)/Pt films at $\mu_0H=14.6$~mT in devices with $D=0.5$, 0.8 and 1.1~$\mu$m. Green solid lines are the amplitude of $\Delta L_{21}$ and $f_{per}/2$ is the half of a signal period. (inset) Dependence of the logarithm of the maximum SW signal amplitude $-\ln(A)$ on the propagation time $\tau$. Solid line is linear fit. (b) Relaxation rate (left scale)/damping(right scale) versus $1/\mu_0M_st$ (symbols). Solid line is a corresponding linear fit.} 
\label{Gamma_vs_t}
\end{figure}

\section{Spin pumping}

A heavy metal acts as an efficient absorber of the spin current generated in an adjacent ferromagnet by the magnetization precession in the uniform ferromagnetic resonance (FMR) mode~\cite{Ando2014,Zhang2015c}. This implies a loss of spin angular momentum in the FM resulting in an increase of damping, which is usually extracted from the enlargement of the FMR linewidth. Naturally, a non-uniform magnetization precession in a form of a spin wave can also pumps the spins out of the FM resulting in an increase of the spin wave relaxation rate ($\Gamma$). To extract the value of $\Gamma$ we use the method described in Refs.~[\onlinecite{Gladii2016a,Collet2017}]. For each film thickness we use devices with different distances $D$ between the antennas in order to follow the amplitude and the period of the signal as function of propagation distance. In Fig.~\ref{Gamma_vs_t} (a) the mutual-inductance spectra measured in the thinnest Py/Pt film with $D$ varying from 0.5 $\mu$m to 1.1 $\mu$m are plotted. As one sees, the period of signal oscillation $f_{per}$ decreases with increasing $D$ and the amplitude of the signal $A$ decays.\footnote{Strictly speaking  $A=\Delta L_{21}/\sqrt{\Delta L_{11}\Delta L_{22}}$  is the maximum amplitude of the mutual-inductance waveform normalized to those of the two self-inductances} This decrease can be written as $A=\exp (-D_{\mathrm{eff}}/L_{\mathrm{att}})$, where $D_{\mathrm{eff}}=D+D_0$ is the effective width of the antenna with a correction term $D_0$ accounting for the finite width of the antenna, and $L_{att}$ is the spin wave attenuation length over which the magnitude of the magnetization precession decay by a factor $e$. Equivalently, the amplitude decay can also be written as $A=\exp(-\Gamma\tau)$, where $\tau=1/f_{per}$ is the spin wave propagation time. Thus, by plotting $-\ln(A)$ as a function of $\tau$ one extracts the value of $\Gamma$ from the corresponding slope, as  shown in the inset of Fig.~\ref{Gamma_vs_t} (a). 

Fig.~\ref{Gamma_vs_t} (b) shows the dependence of the relaxation rate $\Gamma$ (left scale) and the effective damping $\alpha_{eff}$ (right scale) on the inverse of Py thickness. The latter is extracted using the relation $\alpha_{eff}=\Gamma/(\omega_0+\omega_M/2)$, where $\omega_0=\gamma\mu_0H_0$ and $\omega_M=\gamma\mu_0M_s$. In the thinnest Py film adjacent to the Pt layer we measure a damping factor two times larger than in the reference Py/Ti sample, which is due to the spin angular momentum loss at the Py/Pt interface. By increasing the Py thickness $\alpha_{eff}$ decreases obeying a $1/t$ dependence as expected from spin pumping theory~\cite{Tserkovnyak2002}. The damping enhancement is directly linked to spin pumping efficiency via a quantity called the effective spin mixing conductance $g^{\uparrow\downarrow}_{eff}$ as $\alpha=\alpha_0+\alpha^{sp}=\alpha_0+g\mu_B g^{\uparrow\downarrow}_{eff}/4\pi M_st$, where $g$ and $\mu_B$ are the Land\'{e} factor and Bohr magneton, respectively. By using the measured $\alpha$ dependence on $t$, one extracts $g^{\uparrow\downarrow}_{eff}$ from the slope of the corresponding linear fit, which gives $g^{\uparrow\downarrow}_{eff} = 3.2\times 10^{19}$ m$^{-2}$. The fit intercept corresponds to $\alpha_0= 0.008$, which is the typical value of damping factor for Py films~\cite{Haidar_these}. This value is close to the value $\alpha= 0.009$ measured for the Ti/Py(4)/Ti reference film, for which the spin pumping is negligible. We also note that the value of $g^{\uparrow\downarrow}_{eff}$ we measured at $k=7 \mu$m$^{-1}$ is of the same order of magnitude as the one measured with other technique for the $k=0$ FMR mode~\cite{Obstbaum2014}. This is in agreement with the theory~\cite{Tserkovnyak2002} since the effect is not expected to be $k$ dependent, as the SW wavelength remains much larger than the length scales of electron spin transport which are in the nanometer range.


\begin{figure}
\includegraphics[width=7.0cm]{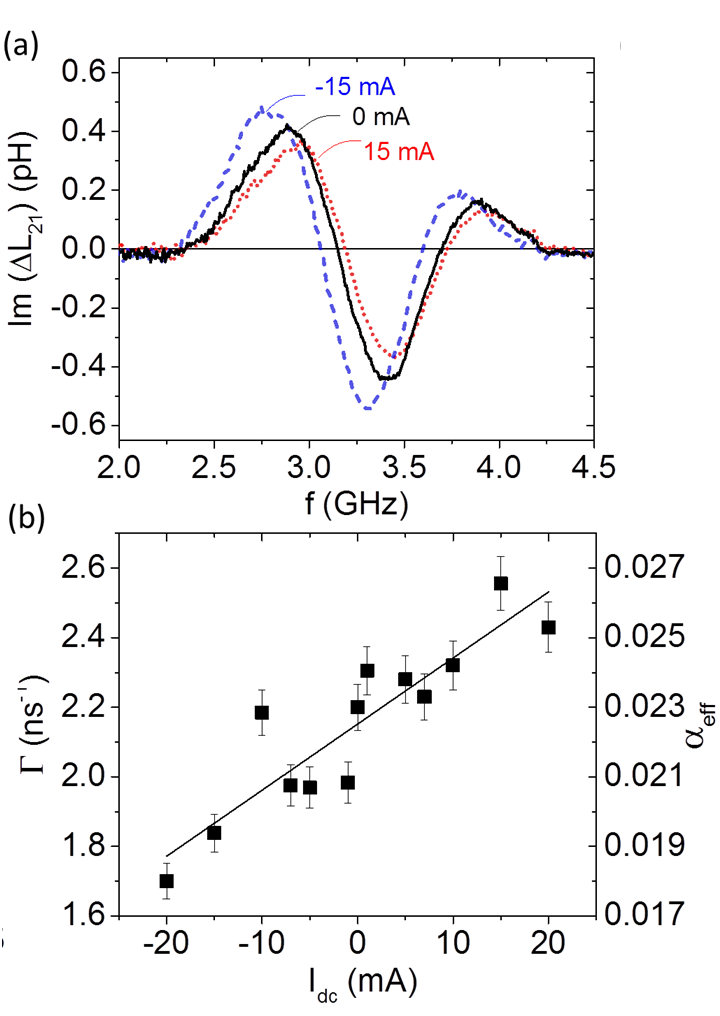}
\caption{(a) Mutual-inductance spectra measured at $\mu_0H=5.8$~mT in a Py(4)/Pt device with $D=0.5 \mu$m at zero current (black line), +15 mA (red dotted line) and -15 mA (blue dashed line). (b) Dependence of the spin wave relaxation rate (left scale) and the effective damping (right scale) as a function of applied $dc$ current. The solid line is a linear fit.} 
\label{Gamma_vs_I}
\end{figure}

\begin{figure}
\includegraphics[width=8.2cm]{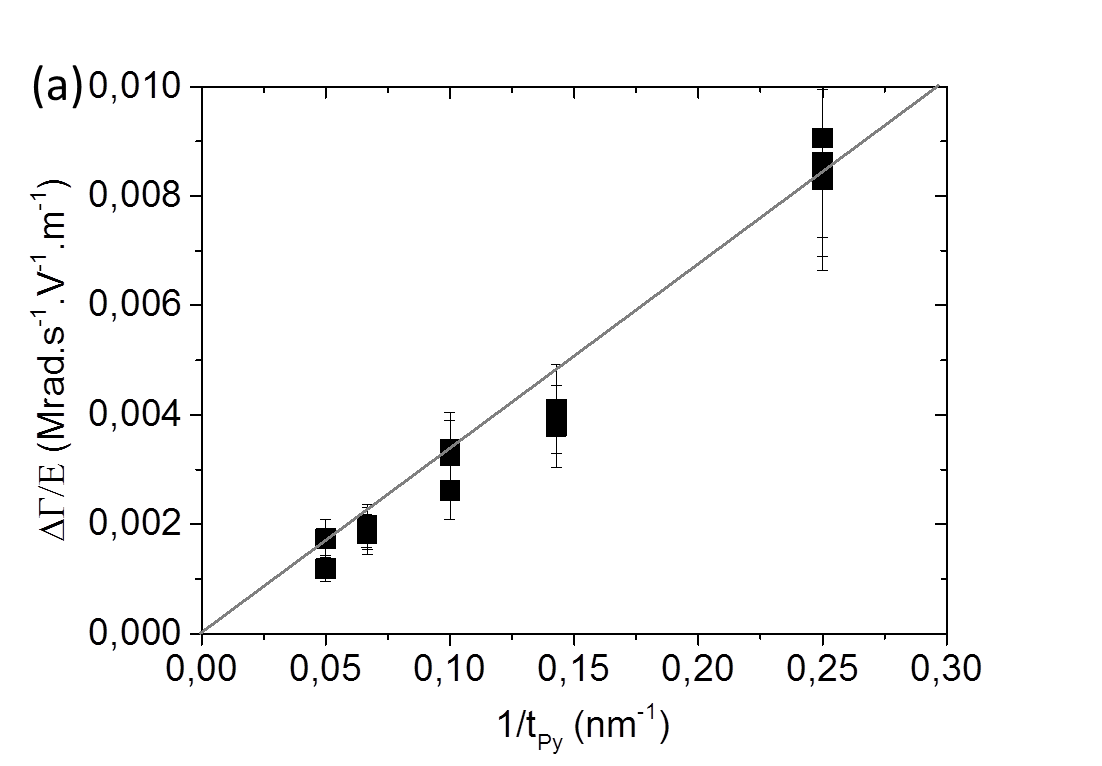}
\includegraphics[width=8.2cm]{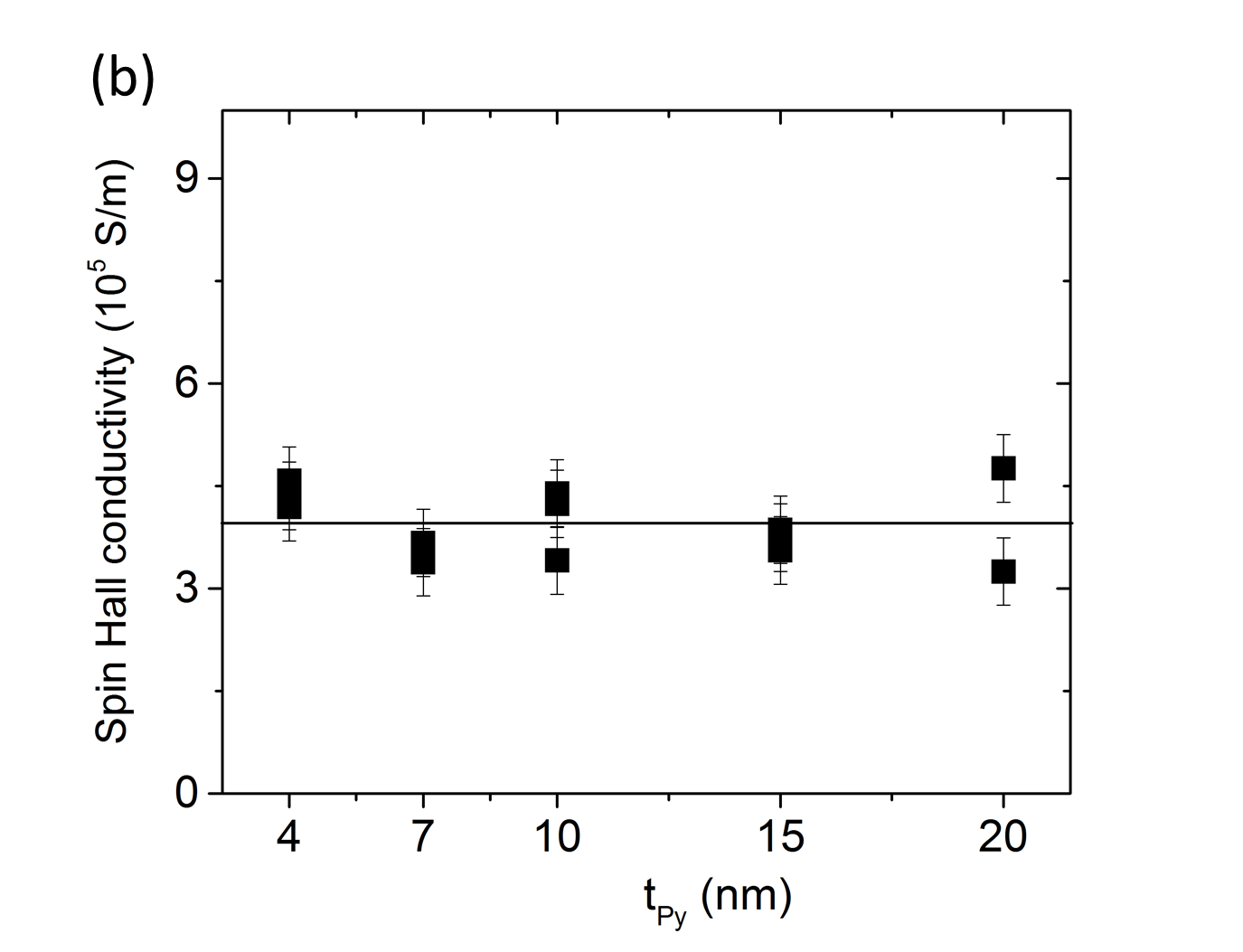}
\caption{ (a)  Measured current-induced change of the spin-wave relaxation rate ($\Delta\Gamma$) per unit of electric field $E$ versus Py thickness (symbols). (b) Dependence of the SHC on the Py thickness. In (a),(b) the lines are guides for the eyes.} 
\label{Gamma_vs_t_at_DC}
\end{figure}

\section{Spin transfer torque induced by spin Hall effect}

To demonstrate the effect of SHE-induced STT on the relaxation rate we inject a \textit{dc} current into the Py/Pt bilayer. The current flowing in the Pt part generates a transverse spin current, which penetrates into the FM and induces a torque on the magnetization. Since the spins are injected either parallel or antiparallel to the local magnetic moment, STT acts as a "damping-like" torque affecting the spin wave relaxation rate. Fig.~\ref{Gamma_vs_I}(a) shows typical mutual-inductance spectra measured at $\pm$15 mA for the film with  4nm of Py at a constant magnetic field. 
One observes a pronounced change of the signal amplitude as well as of the signal frequency with respect to the zero current waveform. The latter is attributed mostly to the Oersted field generated by the current flow in the Pt layer with an additional contribution from the Joule heating, which leads to a slight asymmetry in the frequency shift with respect to the zero current frequency~\cite{Gladii2016a}. The change of the signal amplitude is the direct consequence of SHE-STT. By extracting the relaxation rate by means of the above-mentioned procedure (the comparison of the amplitude and the period of the signal as a function of $D$) for a \textit{dc} current ranging from $-20$ to $20$~mA, we observe a linear increase/decrease of $\Gamma$ ($\alpha_{eff}$) when a positive/negative current is applied [Fig.~\ref{Gamma_vs_I} (b)]. Indeed, depending on the current polarity, the STT will either enhance or reduce the spin wave relaxation rate. For $t=4$~nm the relaxation rate changes by approximately 20$\%$ at $I_{dc}=\pm20$~mA. 

To quantify the SHE-STT for thicker films, we use a simplified method of analysis. Instead of following systematically the mutual-inductance as a function of the distance, we extract directly the current-induced change of relaxation rate ($\Delta\Gamma$) by comparing the maximum amplitude measured at positive ($|\Delta L^+_{ij}|^{max}$) and negative ($|\Delta L^-_{ij}|^{max}$) currents on a single device. Writing  the mutual-inductance as $|\Delta L_{21}|^{max}=\Delta L_0\exp(-\Gamma \tau)$ we can express the SHE-STT-induced variation of relaxation rate as $\Delta\Gamma=-1/\tau \ln(|\Delta L^+_{ij}|^{max}/|\Delta L^-_{ij}|^{max})$, where $\tau$ is deduced from the period of the oscillations. The quantity $\Delta\Gamma$ is then normalized by the electrical field $E$, which can easily be deduced from the voltage drop measured between the two inner pads (Fig.~\ref{device}), in order to compare the effect for different values of $I_{dc}$. Another advantage of this normalization procedure is that it does not require to estimate the current density flowing into the Pt, which would require strong assumptions on the distribution of the current across the thickness of the bilayer. Moreover, this also allows one to extract directly the spin Hall conductivity, a quantity extracted from ab-initio calculations of the intrinsic spin Hall effect (see below).

Fig.~\ref{Gamma_vs_t_at_DC}(a) shows the change of the relaxation rate normalized by the electric field $\Delta\Gamma/E$ as a function of the Py thickness. This quantity rapidly decreases with increasing Py thickness, clearly demonstrating the interfacial nature of the effect. We also observe that the SHE-STT works in the same manner for positive and negative wave vector, \textit{i.e.} it either amplifies or damps the spin wave depending on current polarity but independent on the spin wave propagation direction (not shown). The efficiency of SHE-STT can be quantified by means of the spin Hall conductivity (SHC) $\sigma_{SH}$, which is defined as $\sigma_{SH}=eJ_s/E$, where $J_s$ is the spin current density injected at the Py/Pt interface.\footnote{Here, we define $J_s$ as a number current density, so that the flow of magnetic moment writes $g \mu_B/2 J_s \approx \mu_B/ J_s$. This results in the same convention for $\sigma_{SH}$ as in Ref.\onlinecite{Lowitzer2011}, whereas several other papers use another convention in which the spin Hall conductivity is halved.} By introducing the SHE-STT term in the LLG equation for the surface spin wave configuration one gets a direct relation for the current induced change of the relaxation rate:

\begin{equation}
\Delta \Gamma_{\text{STT}} = \sigma_{SH} \frac{\mu_B}{e M_s t_{\text{Py}}} E.
\label{SHA}
\end{equation}

Using this relation and the values of $\Delta\Gamma/E$ plotted in Fig.~\ref{Gamma_vs_t_at_DC}(a) we extract the spin Hall conductivity. This results in a SHC which is approximately constant within the studied thickness range [Fig.~\ref{Gamma_vs_t_at_DC}(b)] with a value of the order of $4\times10^5$~S/m. In order to translate the spin Hall conductivity in another commonly used quantity, namely the spin Hall angle ($\Theta_{SH}=J_{s}/J_{Pt}$), one can use the average conductivity of our Pt layers $\sigma_{Pt}=4\times10^6$~S/m deduced from the four-wire resistance measurement on different devices,\footnote{These resistance measurements also provide an estimate for the average conductivity of the Py layers $\sigma_{Py}=2.5\times10^6$~S/m and for the fraction of current flowing in Pt layer, found to decrease from 0.85 in the bilayer with 4 nm of Py down to 0.3 in the one with 20 nm of Py} which results in an effective spin Hall angle $\Theta_{SH}=0.1\pm0.02$.

Let us now comment on the spin torque efficiency. The estimated values for the spin Hall conductivity / angle are among the highest reported recently for Pt/Py bilayers~\cite{Obstbaum2016}. Interestingly, our value of SHC is almost the same as the one extracted on single Pt films~\cite{Stamm2017} and those deduced from first principles calculations of the intrinsic SHC of Pt~\cite{Lowitzer2011,Guo2008}, an effect having its origin in the Berry curvature induced by the spin-orbit interaction on the electron bands. It should be noted however that our SHE measurement is an effective one, related to the total spin current penetrating the Py layer: In addition to the spin Hall effect occuring within the Pt layer, it is likely to be influenced also by a number of interface related phenomena, including spin backflow,\cite{Zhang2015b} spin memory loss\cite{Rojas-Sanchez2014}, magnetic proximity effect~\cite{Zhang2015} and interfacial spin current generation~\cite{Amin2018}.

Finally, our measurements show that even if SHE-STT is an interfacial related effect, it can still influence significantly the propagation of spin waves in a relatively thick ferromagnetic layer, which is more suitable for magnonic studies because of its longer spin-wave attenuation length. Combining Fig.~\ref{Gamma_vs_t} (b) and \ref{Gamma_vs_t_at_DC}(a), the relative change of relaxation rate $(\Delta\Gamma/E)/\Gamma$ is found to decrease only by a factor of two between the bilayer with 4nm of Py and that with 20nm, the decrease in SHE-STT being partly compensated by the decrease in the spin pumping contribution to the zero current damping.

\section{Conclusion}
In this paper, we have demonstrated how three different inter-related spin orbit-induced phenomena influence the coherent non-uniform dynamic magnetization of a propagating spin wave in ferromagnetic/heavy metal bilayers. We have shown that the frequencies of two counter-propagating spin waves become non-reciprocal as a result of the combined effect of iDMI and a difference in surface anisotropy,and that the magnitude of both effects can be extracted from the dependence of the frequency non-reciprocity on the magnetic film thickness. The impact of spin orbit coupling on the spin wave relaxation rate was studied by spin pumping and SHE-STT experiments. The analysis of the spin wave relaxation rate with and without $dc$ currents allows us to extract the quantities measuring the strength of both effects, namely the spin mixing conductance and the spin Hall conductivity. The data for ferromagnetic film thickness between 4 and 20 nm could be interpreted using a single set of spin-orbitronic parameters, confirming the consistency of the analysis and the utility of this method to extract the STT induced by the direct spin Hall effect as well as the strength of iDMI in a broad range of experimental conditions.

\section{Acknowledgements}
The authors would like to thank M. Acosta for assistance with sputtering deposition, and the STnano platform and the Labex-NIE (ANR-11-LABX-0058-NIE) for giving us access to nanofabrication equipment. This work was supported by the Agence Nationale de la Recherche (France) under Contract No. ANR-11-BS10-003 (NanoSWITI). O.G. thanks IdeX Unistra (ANR-10-IDEX-0002-02 ) for doctoral funding.

\bibliography{SW_for_spinorbitronics}

\end{document}